\documentclass[twocolumn]{article}
\usepackage[authoryear,round,longnamesfirst]{}
\usepackage{textcomp}
\usepackage{amssymb}
\usepackage{graphicx}
\usepackage{caption}
\usepackage{sectsty}
\usepackage{amsmath}
\usepackage{url}
\sectionfont{\fontsize{10.5}{10}\selectfont}

\newcommand{\twobar}{/\kern-0.2em/}
\title{VR \textquotesingle SPACE OPERA\textquotesingle: MIMETIC SPECTRALISM IN
AN IMMERSIVE STARLIGHT AUDIFICATION SYSTEM}

\author{B. Carey \\ {\small Hochschule f\"{u}r Musik und Theater Hamburg}\\ {\small benedict.carey@hfmt-hamburg.de} \\ \\ {B. Ula\c{s}}\\ {\small \.{I}zmir Turk College Planetarium}\\ {\small bulash@gmail.com}}

\begin{document}
\twocolumn[
\begin{@twocolumnfalse}
\maketitle

\begin{abstract}

This paper describes a system designed as part of an interactive
VR opera, which immerses a real-time composer
and an audience (via a network) in the historical location
of G\"{o}beklitepe, in southern Turkey during an imaginary
scenario set in the Pre-Pottery Neolithic period
(8500-5500~BCE), viewed by some to be the earliest example
of a temple, or observatory. In this scene music is
generated, where the harmonic material is determined
based on observations of light variation from pulsating
stars, that would have theoretically been overhead on the
1st of October 8000~BC at 23:00 and animal calls based
on the reliefs in the temple. Based on the observations
of the stars V465~Per, HD~217860, 16~Lac, BG~CVn and KIC 6382916, frequency collections were derived and applied to the generation of musical
sound and notation sequences within a custom VR
environment using a novel method incorporating spectralist
techniques. Parameters controlling this ‘resynthesis’
can be manipulated by the performer using a Leap Motion
controller and Oculus Rift HMD, yielding both sonic
and visual results in the environment. The final opera is
to be viewed via Google Cardboard and delivered over
the Internet. This entire process aims to pose questions
about real-time composition through time distortion and
invoke a sense of wonder and meaningfulness through a
ritualistic experience.

\end{abstract}
\end{@twocolumnfalse}
]


\section{Introduction}

The system we have developed forms the basis for the
forthcoming opera {\it Motherese}. It immerses a real-time
composer and an audience (via a network) in the historical
location of G\"{o}beklitepe, in southern Turkey during an
imaginary scenario set in the Pre-Pottery Neolithic period
(8500-5500 ~BCE). A description of the networking features
of this package is beyond the scope of this paper,
instead we will concentrate on the virtual staging, sound
resynthesis and sonification aspects of out system.

Rather surprisingly so, FFT analysis, which is so
commonly employed in the composition of spectral music,
originates from a formula designed for rapidly calculating
the elliptical orbits of planetary bodies. This early
version of the DFT is a development attributed to Alexis-
Claude Clairaut in 1754 \cite{c1} but one could look even further back to ancient Babylonian mathematics if the term
\textquotesingle spectral analysis\textquotesingle, which is often used to describe the method by which spectralist composers derive musical material for their compositions, and is sometimes a stand
in for \textquotesingle harmonic analysis\textquotesingle~\cite{c2}. Of course since the term
harmonic analysis already connoted something entirely
different amongst musicologists by the time the French
Spectralist tradition began in the 1970\textquotesingle s, this linguistic
evolution makes sense, despite being a slightly confusing
side effect both of the difficulties of categorization and
the interdisciplinary nature of Spectralism. Mostly the
term spectral analysis is used in an even more narrow
sense when speaking in the context of spectral music
however, to refer to DFT or FFT analysis of audio signals
containing content from within the audible frequency
range (20~Hz and 20.000~Hz) to yield a collection of frequencies
(pitches) and their amplitude (dynamic) variance
over time for a composition. Indeed the stipulation
that spectral analysis produces musical results is a creative
leap of faith that supports the co-option of this process
into the composer\textquotesingle s repertoire of compositional
techniques, and for good reason. Why shouldn’t one look
to mathematics to help build a stronger understanding of
music via recognition of the structural underpinnings of
sound, the very concrete from which this art form emerges?

Yet at the same time why stop at the analysis of
sound to produce frequency collections from which to
derive new harmonies and timbres? FFT analysis is a
tried and tested tool for modelling a musical representation
of a subject, using the program {\it Macaque}\footnote{http\twobar georghajdu.de/6-2/macaque/} in combination
with a SDIF file for example, one can easily track
the movement of a sound spectrum over time such as was
done by G\'{e}rard Grisey through a similar method for his
seminal work {\it Partiels} \cite{c3}, whose methods we will focus
on here. If FFT analysis translates its usefulness so well
from the realm of the cosmos into such a diverse array of
phenomena such as audio signal processing, medical imaging,
image processing, pattern recognition, computational
chemistry, error correcting codes, and spectral
methods for PDEs \cite{c4}, it is perhaps no more worthy a
candidate for the source of frequency based musical inspiration
than any other similar method of observing the
natural world\textquotesingle s many oscillations.

So is the practice of using other algorithmic
methods to interpret natural phenomena any less valid or
useful to the composer? The process of sonification, or audification as it is more commonly known to astronomers,
is fairly widespread due to the pragmatic consequence,
of speeding up time-consuming manual data
analysis. When approaching spectral music composition
in real-time scenarios as is the case in the project presented
in this paper, the speed at which abstractions of these
forms can be realised as sound is paramount to their success
as music of course, but perhaps the most important
aspect is the representation of the entity in music, an entity
which itself does not transmit any sound through the
great vacuum of space, over distances of multiple light
years. It is therefore fairly reasonable to assume that the
usefulness of spectral compositional methods remains,
even if FFT analysis or some tonal system built around
the \textquotesingle natural harmonic series\textquotesingle is removed from this linear
process, and replaced with another algorithm designed to
derive a similar kind of \textquotesingle tonal reservoir\textquotesingle \cite{c2} for our purposes.

The use of starlight audification to create musical
textures has precedent \cite{c5} but has so far not been incorporated
into a real-time spectral composition system. Of
primary relevance to this particular research project is the
clear, discernable embodiment of extra-musical objects
inside of a musical context known as \textquotesingle Mimetic Spectralism\textquotesingle 
\cite{c6}. It may therefore prove no more relevant to us to
base a composition on ‘sound’ itself, once it is abstracted
to the point of mathematical analysis, than on any other
method of analysis of a physical phenomenon, which we
consider a form of embodiment. The apotheosis of sound
as a kind of \textquotesingle living object with a birth, lifetime and death\textquotesingle \cite{c11} as Grisey put it, is not the focus here. Certainly in
the light of careful review by a skilled composer (or just
one with the right software tools), any collection of frequencies
can be stretched through a wide array of aesthetic
extremes, as the practice of spectralism is after all an
impressionistic exercise \cite{c7}.

\section{Spectralism and Belief}

It has been observed that the use of FFT analysis in music
composition may imply an extra-musical dimension to
the piece concerned \cite{c8}. The assertion that what the composer
produces using spectral techniques is music often
comes along with certain presumptions and philosophies
about the nature of sound and music perception. Inevitably,
this extra-musical motivation pushes this music into
the territory of referential expressionism \cite{c9}. One tendency
among proponents of spectralism is to justify their use
of spectral technique by referencing its links to the sciences.
Many proponents of the movement claim that
forms extracted from within sonic events represent a natural
and fundamental order of music as evidenced by the
micro-structure of sound. Despite the scientific origins of
the techniques used in spectral composition, they are of
course not by themselves scientific proofs of \textquotesingle musical
truths\textquotesingle. Instead, it has instead been suggested that when
\textquotesingle new art\textquotesingle is generated from the analysis of \textquotesingle natural\textquotesingle objects,
this indicates naturalism as the philosophical basis
for the art piece concerned \cite{c8}. Extra-musical representation
in spectral music is not always the intention of the
composer, but sometimes it is unavoidable. G\'{e}rard Grisey
exhibits a kind of devotional respect for sound that is almost animistic. Grisey proposed that spectral music
reminds the listener that sound is in fact living.
"Spectralism is not a system. It’s not a system like serial
music or even tonal music. It's an attitude. It considers
sounds, not as dead objects that you can easily and arbitrarily
permutate in all directions, but as being like living
objects with a birth, lifetime and death." (Grisey, 1996)
With this anthropomorphic approach to sound Grisey
displays this reverence towards the source of his compositions
and his muse, sound itself. In the same interview
he mentions that while his music represents a \textquotesingle state of
sound\textquotesingle it is simultaneously a discourse:
“I would tend to divide music very roughly into two categories.
One is music that involves declamation, rhetoric,
language. A music of discourse... The second is music
which is more a state of sound than a discourse... And I
belong to that also. I would put myself in this group.
Maybe I am both, I don't know. But I never think of music
in terms of declamation and rhetoric and language.”
(Grisey, 1996) With this impetus we created a system that explores animism
and discourse through re-synthesis in a ritualistic
setting. We set out to create a music of discourse, which
embodies starlight and animal calls in the music, which is
generated via commonly used spectralist techniques such
as Orchestral resynthesis and ‘spectra as reservoir’
among others. Here we will detail our approach to ‘Mimetic
Spectralism’.

\begin{table*}
\centering
\caption{The parameters for δ Sct type pulsating star
V465 Per. The pulsation parameters $f$, $A$, $\phi$ are taken
from \cite{c10}. Note that ${f_{min}}$ = 13.721 ${A_{max}}$ = 3.5
mmag and ${\phi_{min}}$ = −0.14. The frequency value of $C_4$ ,
261.630 Hz, is taken from \cite{c11}.}
\label{tabchord}
\begin{tabular}{lcccccc}
\hline
$f~(c/d)$&	A(mmag)	&	$\phi$	&	$f^{\prime}$	&	$L$	&	$p$	&	$f^{\prime}\times C_4 (Hz)$		\\
\hline
14.040	&	3.5	&	-0.14	&	1.023	&	1.000	&	0.00	&	267.647			\\
17.208	&	2.3	&	2.05	&	1.254	&	0.657	&	2.19	&	328.084	\\
33.259	&	1.7	&	1.93	&	2.424	&	0.488	&	2.07	&	634.191		\\
13.721	&	1.1	&	3.55	&	1.000	&	0.314	&	3.69	&	261.630		\\
\hline
\end{tabular}
\end{table*}

\section{Theoretical Framework for Starlight Audification}

Pulsating stars simply expand and shrink within their
radius periodically, because of the interior mechanisms
related with their opacity. The observation of this type of
star gives us valuable information about the inner parts of
these stars. The increased opacity inside a pulsating star
helps to produce heat energy that forces the star to expand.
This expansion causes a decrement in the opacity,
which results in a shrinkage. The recurrence of this cycle
makes the pulsating stars a fascinating candidate for astronomical
observation. The observations of the light
variation from a pulsating star results in a wave shaped
variation of light over time (Fig. 1).

\begin{figure}[ht]
\centering
\includegraphics[scale=0.6]{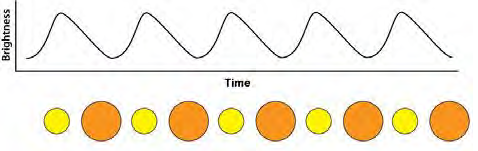}
\caption{The light curve of a pulsating star in comparison
with its radius (Credit: http\twobar www.spaceexploratorium.
com)}
\label{fig1}
\end{figure}

In the case of multiperiodic pulsating stars this wave
shape takes a very complicated form and it can only be decomposed to several sine like variations with certain
frequencies ($f$), amplitudes ($A$) and phase shifts ($\phi$) by
applying frequency analysis. The similarity between observed
-wave shaped- light from pulsating stars and a
superposed simple sound wave is used in converting the
stellar oscillations to audible sounds in our study.

In order to audify the detected oscillation frequencies
of a pulsating star, we used the method described by
\cite{c5}. The author defined three dimensionless parameters
based on the pulsation characteristic of a star: the first
parameter is Relative Frequency, $f^{\prime} = \frac {f_i}{f_{min}}$ , which is the ratio of a given frequency to the minimum frequency
in detected group. The second parameter is the Loudness
Parameter, $L = \frac {A_i}{A_{max}}$ , which is the ratio of the amplitude
value for a given frequency to the maximum amplitude
value among the frequency group. The last parameter,
$p = \frac {\phi_i}{\phi_{min}}$ ,is the Starting Parameter of the signal.
It gives us the difference between the phase shift of a
wave and the minimum phase shift value of the group. A
light variation profile obtained from the star can be converted
to a sound wave by moving the minimum frequency
value to a desired frequency in the audible range and
by keeping the relation between frequencies, amplitudes
and phase shifts. We used five pulsating stars (V465~Per~\cite{c10},
HD~217860~\cite{c14}, 16~Lac~\cite{c15}, BG~CVn~\cite{c16}, KIC~6382916~\cite{c17}) to produce
sounds from the analysis of their observational data. As
an example, we give the pulsation parameters $f$, $A$, $\phi$,
related dimensionless parameters $f^{\prime}$, $L$, $p$ and the result of the multiplication with C4 for one of our stars, V465
Per Table 1.

For the generation of sound waves from these dimensionless
parameters AUDACITY was used. The calculated
relative frequencies for a star was multiplied by
the frequency value of fourth octave C (see Table 1). The
loudness and the starting times are also arranged according
to appropriate values. For instance, when converting
one observed frequency, say 14.040 d-1, of the star V465
Per to audible range we follow these steps: (i) we multiplied
the dimensionless relative frequency by the frequency
value of C4, then we entered the new frequency
value (i.e. 267.647 Hz) as the frequency of a sound wave.(ii) the Loudness parameter (1.000) was entered directly
to the program as the normalized amplitude value. (iii)
The starting time parameter (0.00) was set as the starting
time of the sound in AUDACITY. Since we have 4 observed
frequencies for this star we repeated the process
for each of the 4 frequencies listed in Table 1, therefore,
we obtained 4 different superposed sound waves characterised
by the calculated parameters given in the table.
Finally these sound waves were recorded to a digital
sound file. We hope to expand on this method with
orchestral resynthesis once our system as expanded
beyond the early prototyping stages. Below we detail an
initial implementation utilizing the audio files we created
as described here.

\section{Opera Realisation}

The bulk of the project is realized using the Unity-3d
engine and standard assets, with some additions from the
Unity app store, most notably the Leap Motion Project
Orion Beta, which vastly improves the quality of tracking
possible with the Leap Motion camera in comparison the
previous assets. The standard character controller included
with the Oculus Rift assets was not appropriate due to
our intention to port the system to Google Cardboard\footnote{https\twobar www.google.com/get/cardboard/},
after the initial development done with the Oculus Rift
DK2\footnote{https\twobar www.oculus.com/en-us/dk2/} and Leap Motion\footnote{https\twobar www.leapmotion.com/} camera. Initially there were some problems stemming from the loss of Oculus Rift
DK2 Mac OS X support, these had to be overcome by
porting the project to a Windows~10 development environment.
An important element is the {\it{InstantVR Free}}
asset from {\it{Passer VR}}\footnote{http:\twobar serrarens.nl/passervr/},
which made it possible to port between different VR setups
and platforms relatively easy.

Set design was made easier through importing
freely available photo-scanned models of historical artifacts
or with standard assets as well, saving time on 3d
modelling (Fig.~4.). The majority of the set is actually a
360-degree photograph taken in one of the “temples”; this
was processed into a skybox using the {\it{Panorama to Skybox}}
asset after being edited into a night-like scene in Photoshop.
Some stitching lines are still visible, but they are
mostly obscured with particle systems ranging from fog
to fire and light. The actual characters in the scene are
animated by pre-recorded animations, which are triggered
based on the selections made by the real-time composer.

\begin{figure}[ht]
\centering
\includegraphics[scale=0.6]{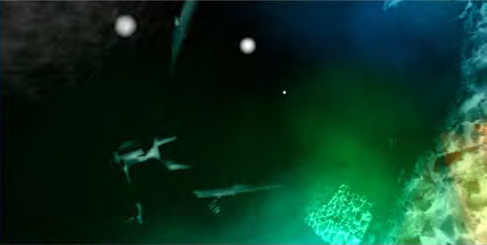}
\caption{The stage from the real-time composers perspective,
3 stars at different levels of luminosity, in the
foreground a fog particle system.}
\label{fig2}
\end{figure}

\section{The User Interface}

\begin{figure}[ht]
\centering
\includegraphics[scale=0.6]{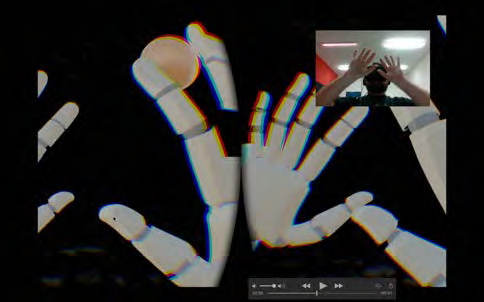}
\caption{A user manipulates the interface with the HMD
mounted Leap Motion Controller.}
\label{fig3}
\end{figure}

The real-time composer controls the playback of
material by selecting single stars with their hand movements
(pinch gesture). Once a particular star is selected,
its partials can be used to manipulate the audio of animal
calls related to the pictograms featured on reliefs at the
G\"{o}beklitepe site. The user is able to manipulate these
sound files from within the VR environment through the
Leap Motion controller and the Unity audio SDK. Visually
the stars themselves increase and decrease in luminosity
in accordance with the relative loudness of each group
(Fig.~2). For example, the real-time composer orients
their hand along the axis of a particular star and their finger
positions affect the amplitude of the sine waves related
to that star. In the case of V465, the user controls the
volumes of 4 sine waves with the degree of extension of
their pinky, ring, middle and index fingers (Fig 3). Using
gesture recognition the user can also open a HUD, populated
by some of the pictograms found throughout the
G\"{o}beklitepe site. Selecting one of these pictograms loads
a sound file related to the particular animal represented
i.e. bison, wild boar, crocodile etc. These sounds can be
used with the {\it{SpectraScore}}\footnote{https:\twobar github.com/benedictcarey/SpectraScorebeta-0.4} Max/MSP abstraction to
generate spectral music, including scores. Various audio
effects allow the user to modify the source sounds in realtime with their movements.


\section{Conclusion}

As this project is in its early stages there is much room
for improvement in terms of the interface and software in
general. Mainly though, the level of latency experienced
between the real-time composers actions and sounding
results needs to be improved to create a smoother \textquotesingle soundbonding\textquotesingle~\cite{c12} effect. In order to achieve this, a new system may have to be created relying on playback of samples
from the audiences HMDs to reduce network strain.
This would hopefully be done with samples of acoustic
instruments such as is currently done with SpectraScore
via MIDI or OSC. Spatialisation would then become a
further layer of complexity, due to the strain of performing
DSP in a smartphone headset.

All in all our success at bringing together techniques
of spectral music composition methods and starlight audification
points at the relative ease through which new
algorithms can be imported into existing algorithmic music
composition frameworks. Since this project was realised
in VR, the exploratory nature of real-time composition
was brought into focus through the use of \textquotesingle source
objects\textquotesingle, that is, \textquotesingle material objects\textquotesingle (Culverwell\footnote{http:\twobar www.oxforddictionaries.com/definition/english/material-object}) that have
been analysed and re-represented in a musical form as
spectral morphemes (representing the physical forms
from which they were derived). This referential expressionist
form of Spectralism creates new possibilities for a
kind of figurative interaction between \textquotesingle Gestalten\textquotesingle that are
otherwise incomparable. Thanks also to an extensive array
of virtualised real-world objects available in online
collections and stores (i.e Sketchfab, Turbosquid, Unity
Asset Store), and the ever increasing documentation surround
the mapping of the sky above the Earth, the possibilities
for sonification with the techniques described here
will continue to grow and increase in relevance for proponents
of the {\it{Gesamtkunstwerk}}.

\begin{figure}[ht]
\centering
\includegraphics[scale=0.23]{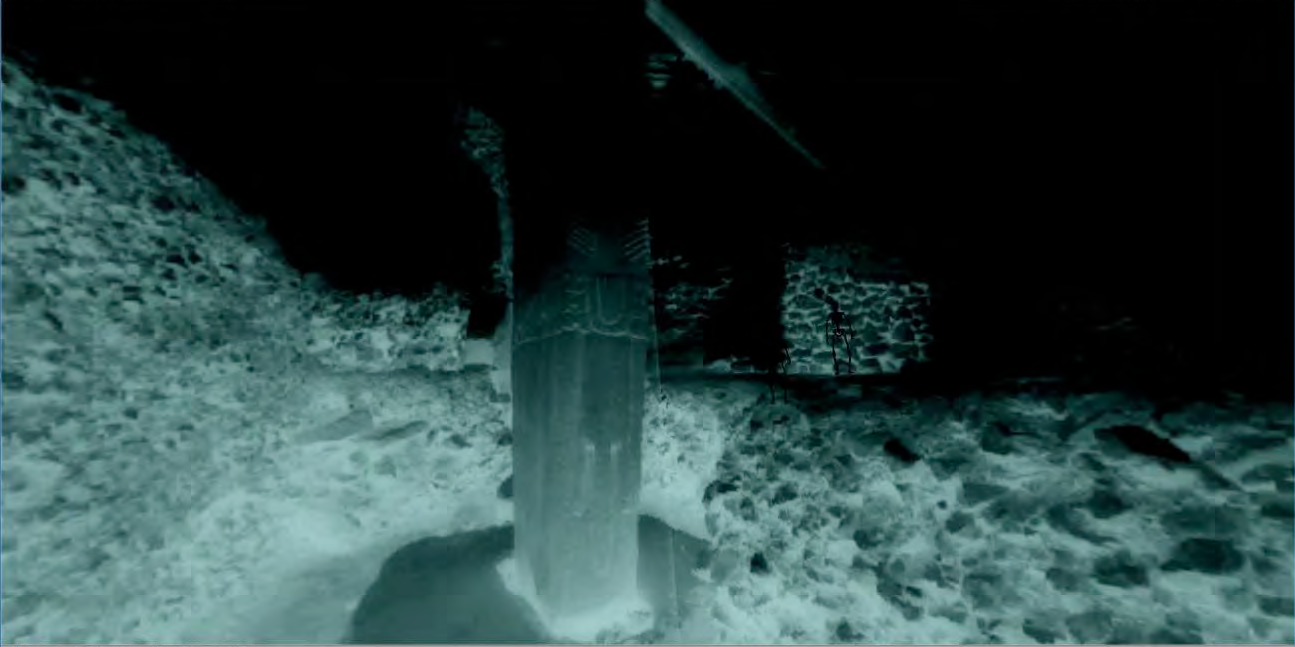}
\caption{A user manipulates the interface with the HMD
mounted Leap Motion Controller.}
\label{fig4}
\end{figure}


\end{document}